\documentstyle[11pt,moriond,epsfig,rotate]{article}

\def\be{\begin{equation}}
\def\ee{\end{equation}}
\def\bea{\begin{eqnarray}}
\def\eea{\end{eqnarray}}

\begin{document}

\begin{titlepage}

\vspace*{1.5cm}

\centerline{\large \bf Theory and phenomenology of generalized parton}
\centerline{\large \bf distributions: a brief overview.}

\vspace{20mm}

\centerline{\bf A.V. Belitsky$^a$, D. M\"uller$^b$}

\vspace{10mm}

\centerline{\it $^a$C.N. Yang Institute for Theoretical Physics}
\centerline{\it State University of New York at Stony Brook}
\centerline{\it NY 11794-3840, Stony Brook, USA}

\vspace{3mm}

\centerline{\it $^b$Fachbereich Physik, Universit\"at Wuppertal}
\centerline{\it D-42097 Wuppertal, Germany}

\vspace{30mm}

\centerline{\bf Abstract}

\vspace{0.8cm}

The generalized parton distributions are non-perturbative objects, which encode
information on long distance dynamics in a number of exclusive processes. They
are hybrids of conventional parton densities, distribution amplitudes and hadron
form factors. We give a brief review of theoretical developments in understanding
of their properties, higher order perturbative effects, power corrections, and
experimental observables where they are accessible.

\vspace{40mm}

\centerline{\it Talk given at the}
\centerline{\it 36th Rencontres de Moriond `QCD and Hadronic Interactions'}
\centerline{\it Les Arcs 1800, March 17-24, 2001}

\end{titlepage}

\vspace*{4cm}
\title{THEORY AND PHENOMENOLOGY OF GENERALIZED PARTON DISTRIBUTIONS:
       A BRIEF OVERVIEW}

\author{ A.V. BELITSKY$^a$, D. M\"ULLER$^b$}

\address{
$^a$C.N.\ Yang ITP, SUNY Stony Brook, NY 11794-3840, Stony Brook, USA \\
$^b$Fachbereich Physik, Universit\"at Wuppertal, D-42097 Wuppertal, Germany }

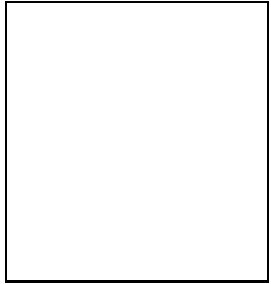
\maketitle\abstracts{
The generalized parton distributions are non-perturbative objects, which encode
information on long distance dynamics in a number of exclusive processes. They
are hybrids of conventional parton densities, distribution amplitudes and hadron
form factors. We give a brief review of theoretical developments in understanding
of their properties, higher order perturbative effects, power corrections, and
experimental observables where they are accessible.}

\section{Exclusive QCD: from form factors to generalized parton
distributions}

The main issue of hadronic physics is the determination of the hadron's
wave function, ${\mit\Psi}$, which encodes a fundamental information on
its structure and dynamics of elementary constituents forming the hadron.
The cleanest indirect access to the former is achieved in lepton-hadron
experiments, which do not suffer from difficulties intrinsic to
hadron-hadron reactions due to initial and final state strong interaction.
Chronologically, these are electromagnetic form factors extracted
from the elastic lepton scattering off hadrons $\ell N \to \ell' N'$, which
provided charge and magnetization distributions within the nucleon. In a
microscopic picture the form factor arises in the parametrization of an
off-forward matrix element of the (local) quark electromagnetic current 
and is related to the wave functions by the Drell-Yan-West overlap formula
\begin{equation}
F (\Delta^2)
=
\langle P_2 | \bar\psi (0) {\mit\Gamma} \psi (0) | P_1 \rangle
\sim
\int dx \int d k_\perp \,
{\mit\Psi}^\ast
\left( x, k_\perp + (1 - x) \Delta_\perp \right)
{\mit\Psi}
\left( x, k_\perp \right) .
\end{equation}
Here $x$ is the longitudinal momentum fraction of the struck parton in the
nucleon and $k_\perp/k_\perp + (1 - x) \Delta_\perp$ being its transverse
momentum before/after the interaction with the probe. ${\mit\Gamma}$ stands
for an appropriate Dirac matrix matching the quantum numbers of the external
states.

The measurements of $F (\Delta^2)$ gave a first insight into the composite
substructure of hadrons, which was unraveled in its full depth in the
pioneering deeply inelastic experiments $\ell N \to \ell' X$ at SLAC. The
Bjorken scaling observed in this reaction, has found an explanation in
the Feynman's naive parton model that expresses the cross section in terms
of a probability density  $f(x)$ to find a partons with momentum fraction
$x$ in a hadron. The emerging probabilistic picture relies on the fact that
the constituents in high energy processes behave as a bunch of noninteracting
quanta at small space-time separations. The rigorous field theoretical basis
is built on the asymptotically free QCD and the use of the factorization
theorems, which give the possibility to separate the contributions responsible
for physics at large and small distances involved in any hard reaction. The
field-theoretical content of the large distance contribution $f (x)$ is given
by the Fourier transform of the forward matrix element of a non-local light-cone
operator ($n^2 = 0$)
\begin{equation}
f (x)
=
\int d \lambda \, {\rm e}^{- i \lambda x P \cdot n}
\langle P | \bar\psi (\lambda n) {\mit\Gamma} \psi (0) | P \rangle
\propto
\int d k_\perp \,
{\mit\Psi}^\ast \left( x, k_\perp \right)
{\mit\Psi} \left( x, k_\perp \right) ,
\end{equation}
where, e.g.\ ${\mit\Gamma} \sim \gamma_+, \gamma_+ \gamma_5, \dots$. Here we
gave also an overlap type representation in terms of hadron wave functions, where
the lowest Fock components have to be kept.

A more direct access to the wave function of a hadron (e.g.\ a meson) is provided
by exclusive production of the latter in the final state, e.g.\ $\gamma^\ast
\gamma^\ast \to \pi^0$. At large momenta of the virtual $\gamma$-quanta
one probes an integral of the lowest Fock component of ${\mit\Psi}$, namely
\begin{equation}
\phi (x) =
\int d k_\perp {\mit\Psi} (x, k_\perp)
=
\int d \lambda \, {\rm e}^{- i \lambda x P \cdot n}
\langle P | \bar\psi (\lambda n) {\mit\Gamma} \psi (0) | 0 \rangle
+ \cdots .
\end{equation}

Recently, we have witnessed the progress in the identification of new
non-perturbative characteristics that contain exhaustive information
on the nucleon wave function. This was triggered by  finding a new
class of parton distributions, the so-called generalized parton
distributions (GPDs), which interpolate between the non-perturbative
functions we have discussed so far. Similarly, to all previous
characteristics, they are defined as Fourier transforms of  non-local
quark/gluon operators sandwiched between the states with different momenta,
e.g.\ for quark fields,
\begin{equation}
\label{GPDs}
A (x, \eta, \Delta^2)
=
\int d \lambda \, {\rm e}^{- i \lambda x (P_1 + P_2)\cdot n}
\langle P_2 |
\bar\psi \left( \lambda n \right)
{\mit\Gamma}
\psi \left( - \lambda n \right)
| P_1 \rangle ,
\end{equation}
and similarly for gluons. Altogether, at leading twist level there are
three quark and three gluon twist-two operators. The parametrization
of e.g.\ the non-local vector current $\bar \psi \gamma_\mu \psi$ involves
two leading twist functions $H$ and $E$ analogous to Dirac and Pauli
form factors. Similarly, in the parity odd sector we have $\widetilde H$ 
and $\widetilde E$, which correspond to the axial-vector and pseudoscalar 
form factors, etc. While being known for a while \cite{DitMueRobGeyHor94},
GPDs have attracted an essential attention only recently, after it 
was realized that they can shed some light onto the spin content of the 
nucleon. Namely, the second moment (Ji's sum rule) of $H + E$ is related 
\cite{Ji96} to the total orbital angular momentum fraction, $J$, carried 
by partons in the nucleon, $\lim_{\Delta \to 0} \int d x \, x (H + E) =
\frac{1}{2} J$.

The cleanest reaction that gives access to GPDs is the deeply virtual Compton
scattering (DVCS), $\gamma^\ast \left( q + \frac{\Delta}{2} \right) N \left(
\frac{P - \Delta}{2} \right) \to \gamma \left( q - \frac{\Delta}{2} \right)
N \left( \frac{P + \Delta}{2} \right)$. In addition to this, they enter
as a soft function in a variety of hard meson production, $\gamma^\ast N \to M
N'$, and diffractive processes. In QCD the Fourier transform (Ft) of the
hadronic part of these processes at large momentum transfer (the answer to
 `How large is it really?' heavily depends on the underlying strong
interaction dynamics and presently this can mostly be judged from experiment only)
is decomposed as
\begin{eqnarray}
\label{DVCSamplitude}
&&\!\!\!\!\!\!\!\!\!\!\!\!\!\!\!\langle N' | T \{ j_\mu (z) j_\nu (0) \} | N \rangle
\stackrel{\rm Ft}{=}
{\cal T}_{\mu \nu}
\int_{- 1}^{1} d x \ C (x, \xi) A (x, \eta)
+
{\cal T}^3_{\mu \nu}
\int_{- 1}^{1} d x \ C^3 (x, \xi) A^3 (x, \eta)
+ {\cal O} ({\cal Q}^{- 2}) , \\
\label{HMPamplitude}
&&\!\!\!\!\!\!\!\!\!\!\!\!\!\!\!\langle N' M | j_\mu (z) | N \rangle
\stackrel{\rm Ft}{=}
{\cal T}_\mu \int_{- 1}^{1} d x \int_0^1 d y \
\phi (y) \alpha_s C (y, x, \xi) A (x, \eta) + {\cal O} ({\cal Q}^{- 1}) ,
\end{eqnarray}
for DVCS \cite{DitMueRobGeyHor94,Ji96,Rad97} and exclusive meson
production \cite{ColFraStr96} amplitudes, respectively. Here $C$ stands
for the hard scattering subprocess, while $A$ and $\phi$ stand for GPDs
and meson distribution amplitudes (DAs), respectively. Here ${\cal T}$ is a
Lorentz tensor. The scaling variables are $\xi = - q^2/ q \cdot P$ as well
as $\eta = q \cdot \Delta/ q \cdot P$ ($\approx - \xi$), and
${\cal Q}^2 = - ( q + \Delta/2 )^2$. In Eq.\ (\ref{DVCSamplitude}) we
have kept the power suppressed contributions, ${\cal T}^3_{\mu \nu}
\propto \Delta_\perp / {\cal Q}$, from twist-three GPDs $A^3$.

\section{Properties and models for GPDs}

In different region of the phase space GPDs share common properties with
conventional parton densities for $|x| > \eta$, and distribution amplitudes
for $|x| < \eta$. GPDs can not be interpreted as densities in general
but rather as interference terms between wave functions of incoming and 
outgoing hadrons \cite{BroDieHwa00}. From the operator definition (\ref{GPDs}) 
it follows that the $j^{\rm th}$ moment of GPDs $A = \{H, E, \widetilde E \}$, 
$A_j (\eta) = \int d x x^{j - 1} A (x, \eta)$, is a polynomial of order $j$ 
in skewedness $\eta$ while for $\widetilde H$ is of $(j - 1)^{\rm st}$ only. 
The sum $H + E$ also obeys the latter property so that the $\eta$ independence 
of the Ji's sum rule \cite{Ji96} is a particular example. To preserve the
polynomiality condition of GPDs the parametrization in terms of spectral,
or double distribution (DD), function reads
\begin{equation}
\label{Htilde}
\left\{
\!\!
\begin{array}{c}
A \\
\widetilde H
\end{array}
\!\!
\right\}
(x, \eta, \Delta^2)
=
\int_{-1}^{1} d y \int_{- 1 + |y|}^{1 - |y|} d z
\delta (y + \eta z - x)
\left\{
\!\!
\begin{array}{c}
x \, F \\
\widetilde F
\end{array}
\!\!
\right\}
(y, z, \Delta^2) .
\end{equation}
While the last definition coincides with the original one introduced by
\cite{DitMueRobGeyHor94,Rad97}, the former \cite{BelMul01a,BelMulKirSch00}
differs from it and respects the polynomiality condition alluded to above.
An alternative solution given in Ref.\ \cite{PolWei99} consists of keeping
for $A$ the representation in terms of DDs of the second line and adding an
independent function $D (x/\eta)$ to it so that this term produces the
missing $\eta^j$-term in the $j^{\rm th}$ moment. The inverse transformation
of DDs in terms of GPDs has been derived in Ref.\ \cite{BelMulKirSch00} and
was found in \cite{Ter01} to be in one-to-one correspondence with the Radon
transformation.

Let us consider a particular example of GPDs. One assumes a factorizable
ansatz of $(x, \eta)$ and $\Delta^2$ dependence: $\widetilde H (x, \eta, 
\Delta^2) = F (\Delta^2) \widetilde H (x, \eta)$, with $\widetilde H (x, \eta)$ 
expressed in terms of a double distribution $\widetilde F (y, z)$ that is 
modeled \cite{Rad99} as
\begin{equation}
\widetilde F (y, z)
= \left\{
\Delta q (y) \theta (y) - \Delta \bar q (- y) \theta (- y)
\right\} \pi (|y|, z) ,
\qquad
\pi (y, z) = \frac{3}{4} \frac{(1 - y)^2 - z^2}{(1 - y)^3} ,
\end{equation}
and obeys the $\widetilde F (y, - z) = \widetilde F (y, z)$ symmetry
\cite{ManPilWei97}. For illustration purposes let us discuss the isotriplet
combination. Assuming an $SU (2)$ symmetric sea $\Delta \bar q^{(3)} (y) =
\Delta \bar u (y) - \Delta \bar d (y) = 0$, we have $\Delta q^{(3)} (y) =
\Delta u_{\rm val} (y) - \Delta d_{\rm val} (y)$ and model it
by a semi-realistic ansatz $\Delta q^{(3)} (y) = g_A \frac{\Gamma (5
- n)}{\Gamma (4) \Gamma(1 - n)} \frac{(1 - y)^3}{y^n}$ and $g_A = 1.26$.
This allows to get a simple analytical representation
\begin{equation}
\label{SimpleModel}
\frac{\widetilde H^{(3)} (x, \eta, \Delta^2)}{g_A F (\Delta^2)}
=
\frac{\left( 1 - \frac{n}{4} \right)}{\eta^3}
\Bigg\{
\theta (x > - \eta)
\left( \frac{x + \eta}{1 + \eta} \right)^{2 - n}
\left( \eta^2 - x + (2 - n) \eta (1 - x)\right)
- (\eta \to - \eta)
\Bigg\},
\end{equation}
where a dipole parametrization $F (\Delta^2) = (1 - \Delta^2/m_A^2)^{-2}$
with $m_A^2 = 0.9 \ {\rm GeV}^2$ is used. One sees that $\widetilde H^{(3)}
( x, \eta)$ vanishes for $x < - \eta$. Other models arise from computations 
in the framework of the bag \cite{JiMelSon97} and the chiral quark soliton 
model \cite{Petetal97,PenPolGoe00} or based on the overlap representation
\cite{BroDieHwa00}.

\section{Power suppressed corrections}

The leading twist-two predictions (\ref{DVCSamplitude}) (with $A^3 = 0$) and
(\ref{HMPamplitude}) are affected by a number of corrections, with power
suppressed effects in ${\cal Q}$ being one of them. They have been addressed
in detail recently for the generalized Compton amplitude. In Ref.\
\cite{AniPirTer00}, using a QCD improved parton picture of Ellis et al.,
it was shown for the example of a scalar target that the twist-three
contributions induce gauge restoring pieces (lacking in the twist-two
approximation) in the Lorentz structure of the Compton amplitude. A
complete OPE-based analysis has been done in \cite{BelMul00b} where
generalized Wandzura-Wilczek relations have been found. The twist-three
GPDs are expressed by these relations in terms of twist-two ones as well as 
interaction dependent antiquark-gluon-quark correlations, $A^3 = W \otimes 
A + \langle \bar \psi G \psi \rangle$. In Refs.\ \cite{KivPolSchTer00,RadWei00} 
it was observed that for the DVCS kinematics, $\eta \approx - \xi$, the 
Wandzura-Wilczek relations alluded to above develop a singularity, however, 
the latter does not show up in the cross section in which the twist-three 
functions enter in a specific combination. In an earlier attempt \cite{BluRob00} 
a generalized Wandzura-Wilczek relation has been derived by neglecting all 
twist-three operators, and is, therefore, erroneous.

Since the cross sections for the reactions in question are peaked at low
momentum transfer, mass corrections  may play a vital role in the 
confrontation of theoretical predictions with experimental measurements. 
As a first step towards this direction we have resummed \cite{BelMul01a} 
the target mass corrections, $\sum_k c_k \left(M^2/{\cal Q}^2\right)^k$, 
stemming from the trace terms of the twist-two operators. The same can 
be done in the twist-three sector and will provide effects of the same 
order. The twist-four sector awaits its unraveling since the renormalon 
analysis demonstrated a potentially sizable effect coming from the latter 
already for the moderate momentum transfer \cite{BelSch98}.

\section{Perturbative corrections}

Leading order calculations in QCD are known to be unreliable and they 
are strongly affected by perturbative corrections. For the cases at 
hand, the coefficient function is a series in coupling
\begin{equation}
C = C_0 + \frac{\alpha_s}{2 \pi} C_1 + \cdots .
\end{equation}
The one-loop correction $C_1$ to the handbag approximation of the
generalized Compton scattering amplitudes have been derived in
\cite{JiOsb97,BelMul97,ManPilSteVanWei97}. For the hard meson
production with quark dominated parton subprocess, $C_1$ has been
recently extracted  in Ref.\ \cite{BelMul01b} making use of the known
result for the pion form factor. The analyses have shown that they
can produce a very large modification of the leading order (LO)
predictions \cite{BelMulNieSch99,BelMul01b}.

Yet another source of corrections stems from the evolution of GPDs
(and DAs), which is governed (at twist-two level) by the equation
(with $\eta = 1$ for DAs)
\begin{equation}
\label{EvolutionEq}
\frac{d}{d \ln {\cal Q}^2} A (x, \eta)
= \int_{- 1}^{1} d y
K \left( x, y, \eta \right) A (y, \eta),
\quad
\mbox{with}
\quad
K = \frac{\alpha_s}{2 \pi} K_0
+
\left( \frac{\alpha_s}{2 \pi} \right)^2 K_1
+ \dots .
\end{equation}
The one loop kernels for all channels have been computed by Lipatov
et al.\ in Ref.\ \cite{BukFroKurLip85} and recalculated by a number
of groups in the last few years of XX$^{\rm th}$ century (see \cite{Ji96}
for a complete list of references). The two-loop quantities have become
available first in the form of local anomalous dimensions from Refs.\
\cite{Mul94,BelMul98,BelMul00a} and later have been converted into the
momentum fraction form in \cite{BelMulFre99} by means of an extensive
use of conformal \cite{BelMul98} and supersymmetric \cite{BelMulSch98}
Ward identities. Note that the flavour nonsinglet kernel has been known
for quite a while from \cite{DitRad84,Sar84,MikhRad85}.

The evolution effects have been studied by methods based on the orthogonal
polynomial reconstruction \cite{BelGeyMulSch97,ManPilWei97}, extended in
\cite{BelMulNieSch98} to NLO accuracy, and the direct numerical integration
\cite{MusRad99}. The resulting two-loop effects have been found to alter the
LO evolution in a percentage range.

\section{Observables}

In the electroproduction process of a real photon the DVCS signal is
strongly contaminated by the Bethe-Heitler (BH) process, $\sigma
\propto |{\rm DVCS} + {\rm BH}|^2$. However, individual terms in this
sum are distinguished by peculiar dependencies on the lepton
charge, lepton $\lambda$ and hadron spin, and the azimuthal angle
$\varphi$ of the outgoing photon. The best quantity to access the GPDs
is, of course, the DVCS-BH interference since the GPDs enter
linearly in it. Its precise experimental extraction is only possible
on machines having lepton beams of both charges, $\sigma^+ - \sigma^-
\propto$ DVCS-BH.  This allows a clean separation of Fourier
components w.r.t.\ azimuthal angle, DVCS-BH $\propto \sum_m [c_m \cos ( m
\varphi) + \lambda s_m \sin (m \varphi)]$. For a lepton beam with a
given charge, other observables (like diverse spin asymmetries, which
isolate the DVCS-BH interference at leading power in $1/{\cal Q}$-expansion)
will be contaminated by (sizable) power suppressed (higher twist)
effects. For instance, in the cross section $\sigma$, the coefficient
of $\cos/\sin (\varphi)$ stemming from the twist-two DVCS-BH will get
corrected by $|{\rm DVCS}|^2$ at twist-three level, etc. Note however,
that the twist-three effects do not induce the same angular dependence
in DVCS-BH and $|{\rm DVCS}|^2$, separately, which is already generated
by leading twist contributions, i.e.\ in DVCS-BH they produce $\cos/\sin
(2 \varphi)$ only (for a numerical study see \cite{KivPolVan00}), so that
the coefficient of $\cos/ \sin (\varphi)$ will be modified at twist-four
level \cite{BelMulKirSch00}. The experimental isolation of different Fourier
components can be done by forming appropriate azimuthal asymmetries or
weighting data with corresponding angular functions $\cos/\sin (m \varphi)$.
As distinguished from DIS, in DVCS one can access tensor gluons, which are
not contaminated by quarks at twist-two. They show up at one-loop order
\cite{HooJi98,BelMul00a} and generate a specific $\cos/ \sin (3 \varphi)$
azimuthal angle dependence in the cross section \cite{BelMul00a,Die01}.
Provided we have separated DVCS-BH interference alone, the single spin asymmetries
make it possible to extract  the imaginary part of the DVCS amplitude
and thus give access to the shape (at LO in $\alpha_s$ in complete analogy to
DIS) of GPDs on the diagonal $x = \xi$, e.g.\ \cite{DieGouPirRal97,BelMulNieSch00}
\begin{equation}
\frac{d \sigma^{\leftarrow} - d \sigma^{\rightarrow}}{
d x_{\rm B} d {\cal Q}^2 d |\Delta^2| d \varphi}
= \frac{\alpha_{\rm em}^3}{\pi} \frac{y^2 (2 - x_{\rm B})(2 - y)}{
\sqrt{1 - y} {\cal Q}^5 \Delta^2}
|\Delta_\perp| \sin \varphi \, {\rm Im} \!
\left\{
F_1 {\cal H} + \frac{x_{\rm B}}{2 - x_{\rm B}} G_M \widetilde {\cal H}
- \frac{\Delta^2}{4 M^2} F_2 {\cal E}
\right\} ,
\end{equation}
where ${\cal A} = C \otimes A$, see Eq.\ (\ref{DVCSamplitude}), $F_{1}$,
$F_2$ and $G_M = F_1 + F_2$ are Dirac, Pauli and magnetic Sachs form
factors, and the transverse momentum squared is $\Delta_\perp^2 = \left(
(1 - x_{\rm B}) \Delta^2 + {x_{\rm B}}^2 M^2 \right)/(1 - x_{\rm B}/2)^2$.

In hard exclusive meson production one measures products of off-forward
amplitudes, e.g.\ the asymmetry of the $\pi^+$-production off a
transversely polarized proton reads \cite{BelMul01b,FraPobPolStr99}
\begin{equation}
\frac{d \sigma_{\Uparrow} - d \sigma_{\Downarrow}}{ d |\Delta^2| d \varphi}
= - \alpha_{\rm em} \frac{4 \pi}{9} \frac{f_\pi^2}{{\cal Q}^6}
\frac{x_{\rm B}^3}{2 - x_{\rm B}}
\frac{|\Delta_\perp|}{M}
\sin \varphi \, {\rm Im} \!
\left\{
\widetilde {\cal H}^{(3) \ast} \, \widetilde {\cal E}^{(3)}
\right\} ,
\end{equation}
with $( \widetilde {\cal H}, \widetilde {\cal E} )= \phi \otimes C \otimes
( \widetilde H, \widetilde E )$, see Eq.\ (\ref{HMPamplitude}). Cross sections
for production of other meson species can be found in
\cite{ManPilWei97,FraPobPolStr99,ManPilWei98,VanGuiGui98,ManPilRad99,VanGuiGui99}.
Note that quark transversity GPD does not show up in the meson production due
to preservation of the chiral symmetry in perturbation theory \cite{DiePirGou98}.
The current experimental situation is discussed in \cite{Ama00} and \cite{Sab01}
for HERMES and JLab settings, respectively.

\section*{Acknowledgments}

One of us (A.B.) is deeply indebted to G. Korchemsky and J. Tr\^an Thanh V\^an
for giving him an opportunity to participate in this stimulating event and financial
support within the TMR network. He thanks the Institute for Nuclear Theory at the
University of Washington for its hospitality and the Department of Energy for
partial support during the completion of this work.

\end{document}